\documentclass[useAMS,usenatbib]{mnras}
\usepackage{epsfig, bm, times, aas_macros}
\usepackage[english]{babel}
\usepackage{amsmath}
\usepackage{amssymb}
\usepackage{graphicx}
\usepackage{xcolor}

\newcommand{\src}{XTE J1814--338}
\newcommand{\msun}{$~M_{\odot}$}

\newcommand{\nacc}{N_\mathrm{acc}}
\newcommand{\nbur}{N_\mathrm{bur}}
\newcommand{\racc}{r_\mathrm{acc}}
\newcommand{\rbur}{r_\mathrm{bur}}
\newcommand{\dpm}{\Delta \varphi_m}
\newcommand{\dpb}{\Delta \varphi_b}
\newcommand{\dpr}{\Delta \varphi_u}

\title[Burst Oscillation Drifts]{The Peculiar Behavior of Burst Oscillations in the Accreting Millisecond X-ray Pulsar XTE J1814--338}
\author[Cavecchi and Patruno]{Yuri Cavecchi$^1$\thanks{ycavecchi@astro.unam.mx} and Alessandro Patruno$^2$\\
$^1$ Instituto de Astronom\'ia, Universidad Nacional Aut\'onoma de M\'exico, Ciudad de M\'exico, CDMX 04510, Mexico\\
$^2$ Institute of Space Sciences (IEEC--CSIC), Carrer de Can Magrans s/n, E-08193 Barcelona, Spain}

\pagerange{\pageref{firstpage}--\pageref{lastpage}} \pubyear{}

\begin{document}
\maketitle
\label{firstpage}

\begin{abstract}
Accreting millisecond X-ray pulsars (AMXPs) show burst oscillations during thermonuclear explosions of the accreted plasma which are markedly different from those observed in non-pulsating low mass X-ray binaries. The AMXP XTE J1814--338 is known for having burst oscillations that are phase locked (constant phase difference) and coincident with the accretion powered pulsations during all its thermonuclear bursts but the last one. In this work we use a coherent timing analysis to investigate this phenomenon in more detail and with higher time resolution than was done in the past. We confirm that the burst oscillation phases are, \textit{on average}, phase locked to the accretion powered pulsations. However, they also display moderate ($\lesssim 0.1$ cycles) drifts during each individual burst, showing a repeating pattern that is consistently observed according to the thermonuclear burst phase (rise, peak, tail). Despite the existence of these drifting patterns, the burst oscillation phases somehow are able to average out at almost the exact position of the accretion powered pulsations. We provide a kinematic description of the phenomenon and review the existing models in the literature. The phenomenon remains without a clear explanation, but we can place important constraints on the thermonuclear burst mechanism. In particular, the observations imply that the ignition point of the thermonuclear burst occurs close to the foot of the accretion column. We speculate that the burning fluid expands in a backward tilted accretion column trapped by the magnetic field, while at the same time the burning flame covers the surface.
\end{abstract}

\section{Introduction}
\label{sec:intro}

The accreting millisecond X-ray pulsar (AMXP, see \citealt{patrunowatts2021} for a review) XTE J1814--338 is a 4.3 hr binary composed by a neutron star spinning at 314 Hz and a main sequence companion of minimum mass of about 0.2\msun. The system is a transient and so far has shown an outburst in 2003~\citep{mar03} lasting for ${\approx} 50$ days, with an earlier outburst tentatively identified in 1984~\citep{wij03c}. 

During its 2003 outburst, the neutron star has shown 28 thermonuclear bursts (Type I bursts) which were studied in detail by~\citet{str03,wat05,bha05} and \citet{wat06}, and were identified as sub-Eddington bursts, with the exception of the last one that was brighter than the others and showed possible signs of photospheric radius expansion \citep{str03}. Burst oscillations \citep[BOs, see][for reviews]{wat12, bhatta2021} were observed in all bursts and a peculiarity of the first 27 bursts was that the corresponding BOs were coincident in phase~\citep{str03} and phase locked to the accretion powered pulsations (APPs) with extraordinary precision: within ${\lesssim} 3^{\circ}$~\citep{wat08}. 
The accretion powered pulse phases were observed to fluctuate over time, following a tight anti-correlation with the X-ray flux variations~\citep{has11}, which is a well known phenomenon observed in several other AMXPs~\citep{pat09f}. The phase locking of the BOs implied that any burst asymmetry at the origin of the burst oscillations had to be coupled to the hot spot location. This might not necessarily be coincident with the magnetic field poles, since these are assumed to be fixed on the surface of the neutron star (at least over the timescale of an outburst) whereas the hot-spot is observed wandering on the neutron star surface~\citep{pap07,wat08,pat09d,has11}.

An open question is how, in accreting neutron stars, the nuclear burning occurring during a Type I burst is related to the appearance of burst oscillations. Indeed there is no clear mechanism that has been conclusively identified, so far, at the origin of these oscillations. Furthermore, BOs have different properties according to the type of source. For example, it is well known that BOs in AMXPs behave differently from other low mass X-ray binaries where no accretion powered pulsations are seen (for reviews, see \citealt{gal08,wat12,patrunowatts2021}). In persistent AMXPs the BOs have harmonics, even though with smaller amplitude than the APPs. BOs from persistent AMXPs appear in all bursts, while in other sources (including intermittent AMXPs) they may not appear, or not be present throughout the entire burst duration. In the AMXPs XTE J1814--338 and IGR J17480-2446 the BOs track the APPs very closely~\citep{wat08, cav11}, while in SAX J1808 the frequency of the oscillations is varying during the burst rise, and tracks the APP pulse frequency in the burst tail~\citep{chak03}. This has led some authors to suggest that the BO mechanism in AMXPs must be related to, or at least influenced by, the presence of a dynamically important magnetic field~\citep{Lovelace07}.

In this paper we present a new analysis of the BOs in \src{} and we show that, during the thermonuclear bursts, the phase of the BOs shows a distinctive pattern that is consistent across the bursts. In particular we show that the location of the burst oscillations wanders, in a systematic way, around the location where the accretion powered pulses are generated, while at the same time still maintaining a tight average phase locking. This phenomenon was previously observed in a couple of bursts of XTE J1814--338 \citep{str03}, but not recognized as a systematic pattern. We suggest that the magnetic field of the pulsar is responsible for the observed pattern, although we are unable to identify a specific mechanism able to explain all observations. 

\section{X-Ray Observations}
\label{sec:obs}

The X-ray observations and data reduction procedure used in this work are similar to what has been presented in \citet{wat08} with the difference that we use a higher time resolution for the lightcurve segments that we use for epoch folding to investigate the properties of burst oscillations. In particular, while \citet{wat08} folded the entire burst segment into one pulse profile, we use a higher time resolution by creating smaller and independent segments (i.e. non overlapping). We use all high resolution Event 122$\rm\,\mu s$ and GoodXenon 1$\rm\,\mu s$ data from the \textit{Rossi X-Ray Timing Explorer} (\textit{RXTE}) collected by the Proportional Counter Array in the absolute channels 5 to 37 (see \citealt{Jahoda06}). This corresponds to an energy range of approximately 2.5-16 keV (which varies slightly between different Proportional Counter Units, PCUs). We then select the burst intervals, defined as the locations in the lightcurve where the X-ray flux becomes twice the pre and post-burst level.

To measure the pulsations, both during the bursts and outside them, we divide the data into independent segments. For each segment we fold the photon times of arrival into a unique profile of the rotational period according to the pulsar ephemeris. The ephemeris used to fold the data consist of a circular Keplerian orbit and a constant pulse frequency, as reported in \citet{wat08}. We then fit a sinusoid at the pulsar frequency (which is known to many significant digits) minimizing the residuals squared to determine the pulse amplitude and phase (or time of arrival). Minimizing the $\chi^2$ is equivalent to taking a Fourier transform and therefore the resulting amplitude and phase of the oscillations are independent of the presence of other harmonics. We refer to \citet{pat10,hart08,taylor92,scargle82} for further details.

We fold the data outside the burst intervals in segments of varying lengths. First we use approximately 500 second segments and then we repeat the procedure by using the whole duration of the satellite orbit. For the BOs we use a higher time resolution within the bursts, by folding segments of significantly shorter length (between 5 and 30 seconds). We then select 20 s as the optimal choice for the high time resolution, given the existing counting statistics. This is possible because the high flux during the bursts allows the creation of shorter pulse profiles with similar counting statistics. We then select significant pulsations defined as those where the ratio between the pulse amplitude and its 1-$\sigma$ statistical error is larger than 3.3, in order to have less than 1 false positive pulse detection in our sample. We consider only the pulse phases of the fundamental (for both the burst and accretion powered pulses) since the signal-to-noise (S/N) ratio for the second harmonic is not high enough for a sufficiently large number of detections in short 20 s segments during the bursts. We stress that all 20 s long segments are independent of each other (i.e., no sliding window). Both burst oscillations and accretion powered pulse phases are defined such that a positive phase residual corresponds to a delay in the time of arrival with respect to the model, whereas the negative sign means the pulsations lead the model prediction.

\section{Results}
\label{sec:results}

\begin{figure}
\centering
\includegraphics[width=0.48\textwidth]{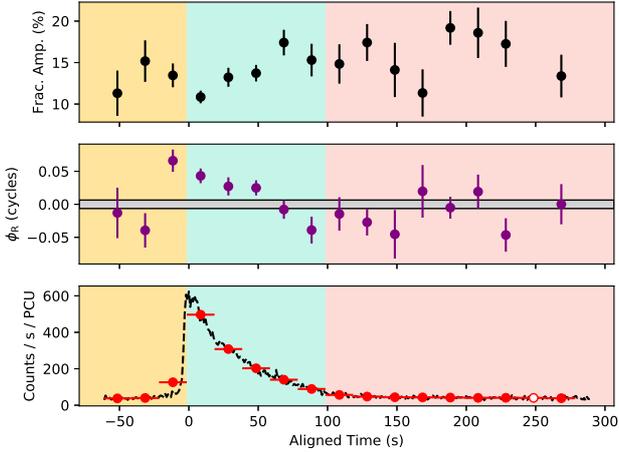}
\caption{Burst nr. 27. The top panel shows the fractional amplitude of the fundamental frequency of the burst oscillations (only significant detections, with their 1-sigma error bar). The middle panel refers to the BO phase residuals with respect to the APP phase (only significant detections, with their 1-sigma error bar) and the bottom panel is the lightcurve of the burst. Here the black dashed line is the 1 second binned lightcurve for reference, while the dots indicate the bin centers used for the analysis (empty if the BOs were not detected significantly). The yellow shaded region is the so-called ``drift'', the blue shaded one is the ``pullback'' and the red shaded one is the ``plateau''. The horizontal gray bar is the APP 68\% confidence interval right before the onset of the burst (centred at zero since we subtract the APP phase).} 
\label{fig:bos27}
\end{figure}

We begin by inspecting the BO phases with respect to a constant spin frequency solution ($\nu_s=314.35610870$ Hz), in a way similar to what was done in \citet{wat08}. In Figure~\ref{fig:bos27} we plot the burst oscillations phase residuals of burst nr. 27, in units of cycles, subtracting the location of the accretion powered pulses (APPs) observed in the segment preceding the onset of the burst (we use the phase measured in the last 500 seconds for more precision). We begin with burst nr. 27 because it is one of the bursts where the effect we are about to discuss is most evident. The APP 68\% confidence interval is highlighted with a gray horizontal bar.

In the first two points in the figure, the counts are still partially contaminated by the accretion powered emission. However, at the third point in the plot, the counts are heavily dominated by the burst emission and therefore the BO phases suffer by very little contamination from the APPs. 
The bias in the burst oscillation phases introduced by any residual accretion powered pulsations can be estimated by knowing $\Delta\varphi_m$, which is the \emph{measured} offset between the measured APP and BO phases \citep{wat08}:
\begin{equation}
\tan{\dpm} = \frac{\sin(\dpm - \dpb)}{\left[\nacc \racc / \nbur \rbur \right] + \cos(\dpm - \dpb)}.
\label{offset}
\end{equation}
Here $N_\mathrm{acc}$ and $N_\mathrm{bur}$ are the number of accretion and burst photons in the data used to obtain the pulse profile, while $r_\mathrm{acc}$ and $r_\mathrm{bur}$ are the \emph{real} APP and BO fractional amplitudes and $\Delta\varphi_b$ is the burst oscillation phase bias, i.e., the offset between the observed and true BO phases. This expression assumes that the measured pulse profiles result from the addition of two offset sinusoidal pulses (one from the APPs and one from the BOs). Simplifying the previous equation we find: 
\begin{equation}
    \dpb = -\sin^{-1}\left[\,R\,\sin(\dpm)\right]
\end{equation}
where $R=\left(\nacc\racc\right)/\left(\nbur\rbur\right)$. The phase residuals shown in this paper refer to the observed BO phases, not the unbiased ones, therefore it is important to keep in mind that a correction $\Delta\varphi_b$ needs to be applied to each point reported. Note also that we are considering only the fundamental frequency of the pulsations, therefore pulse shape changes due to the relative variation of the harmonic amplitudes do not affect our results. 

It can be seen that taking into account a possible contamination by the accretion powered pulsations does actually increase the significance of the offsets we measure. For example, in the yellow and red regions of the figures we can take an indicative value of $\Delta\varphi_m = \pm 0.05$. Since $N_{\mathrm{bur}} = 2 N_\mathrm{acc}$, $R = 0.5 \racc / \rbur$. For the rms of the accretion pulsations we can take an average value of $\racc = 10\%$ \citep{wat05}. Let's consider two cases: $\rbur = 15\%$, i.e. the value in our observations, and $\rbur = 1.6\%$, i.e. close to the minimum allowed by Equation \ref{offset}. In the first case $R=0.3$, which gives an unbiased phase offset of $\dpr = \dpm - \dpb = \pm 0.066$. In the second case $R=3.125$ and we obtain $\dpr = \pm 0.258$. Basically, by increasing the contribution of accretion effects, the bias towards 0 increases and therefore the real offset has to be larger\footnote{Clearly, if there is no contamination $R=0$ and the correction is $0$.}.

In the blue region we can take as an indicative value $\nbur = 5 \nacc$. If we repeat the calculations for the two cases $\rbur=15\%$ and $\rbur=1.6\%$, we obtain $\dpr = \pm 0.057$ and $\dpr = \pm 0.11$. In this case the effect is lower because the counts of the bursts are higher and therefore increase the weight of the burst oscillations on the measurement of the offset $\dpm$. In any case, across the whole burst duration any contamination from accretion enhances the trend we are about to describe. Keeping that in mind, we prefer not to explicitly subtract this contamination because it is difficult to estimate precisely and any timing conclusion based on such a procedure would be model dependent.

From Figure~\ref{fig:bos27} it can be seen that at the burst onset (first two points of the lightcurve) the BOs may be leading the APPs by about 0 - 0.05 cycles. This is seen also in a few other bursts. In these bins the burst flux is low, meaning that it is at most comparable to the pre-burst flux level: this would imply that, \textit{if there is a BO contribution to the phase}, the real BO offset would be even larger. However, since the burst flux is this low, it is perhaps more likely that there is no BO signal in those bins and that we are only detecting APP fluctuations, which are due to timing noise. Therefore, we report this feature, but stress that it is only tentative and may not be due to the BO phenomenology. From the third point on, the APP phase contamination rapidly drops, since the flux of the burst is quickly rising, and the BO phase variations cannot be ascribed to the APP anymore.

As soon as the burst flux increases, the BOs drift by about 0.1 cycles (we refer to this as ``the drift'', highlighted in yellow in Figure~\ref{fig:bos27}), with the final BO phases now lagging the APPs. After this, the BO phases move back towards their original position on a timescale of $\approx150$--$200$ seconds (we call this ``the pullback'', highlighted in blue in Figure~\ref{fig:bos27}), then reach a plateau, possibly turning back by another 0.05 cycles during the final $\approx 100$ seconds of the burst tail (we call this phase ``the plateau'', highlighted in light red in Figure~\ref{fig:bos27}). 

\begin{figure}
\centering
\includegraphics[width=0.48\textwidth]{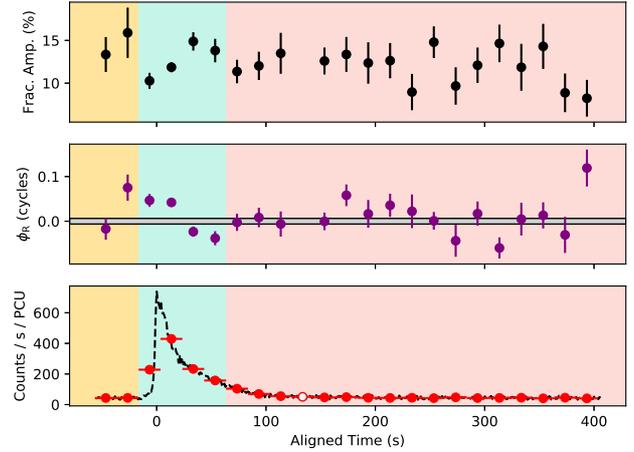}
\caption{Burst nr. 11. For a description of the three panels see caption in Figure~\ref{fig:bos27}.} 
\label{fig:bos11}
\end{figure}

In Figure~\ref{fig:bos11} we show another similar case (burst nr. 11). This time the drift is perhaps already seen early in the burst, with the BO phases already lagging the APPs at the very early stages of the bust rise, even though the error bar on the phase is somewhat large and this could be again a case of APP wandering due to timing noise. However, by the third point the BOs show that they have clearly drifted. After the drift, the BO phases pullback by approximately 0.1 cycle, leading again the APPs and then again they stabilize in a plateau approximately matching the location of the APPs. The overall duration of the drift and pullback is approximately 100 seconds.

\begin{figure*}
\centering
\includegraphics[width=0.5\textwidth]{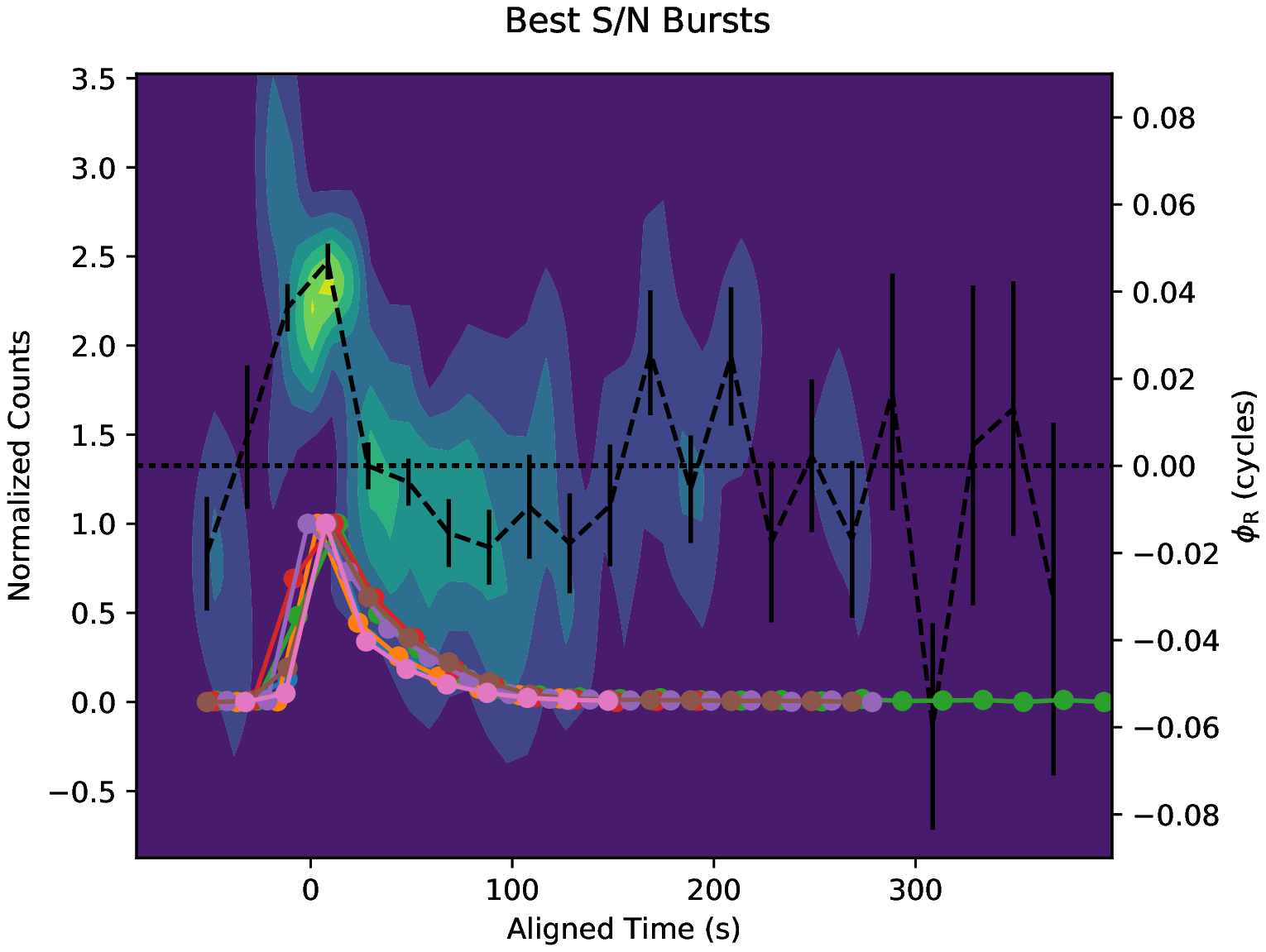}\hspace{\stretch{1}}%
\includegraphics[width=0.5\textwidth]{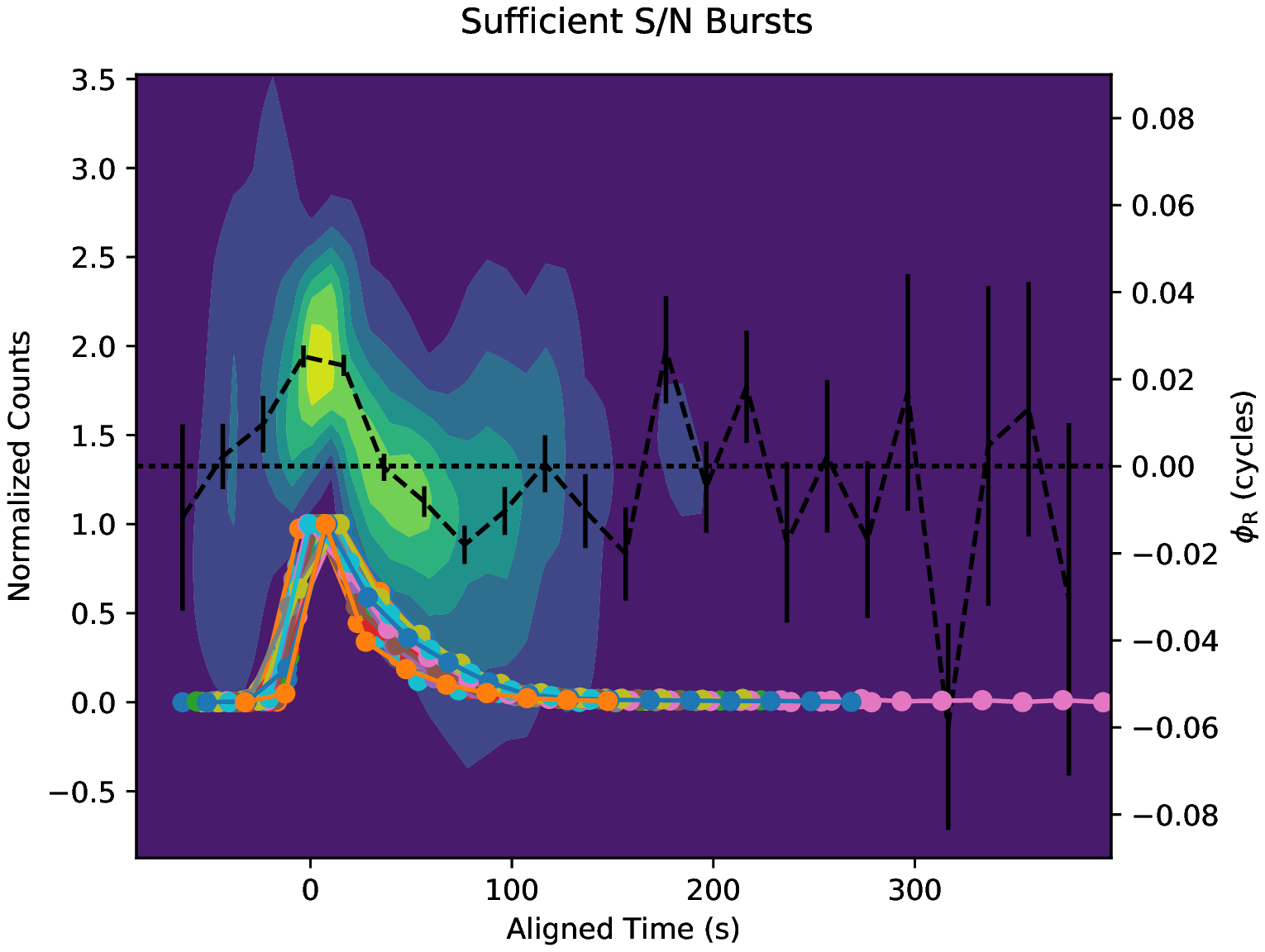}
\caption{The trend of the BO evolution during the bursts for the sample with enough S/N. During the burst rise, the BOs suddenly drift, lagging the reference phase (the APP phase before the bursts) around the burst peak, and then pullback again towards the reference phase during the tail. In the lower part we overplot the burst lightcurves for reference (scale on the left y axis). They are aligned with respect to their peak and the intensity is normalized between $0$ and $1$. This is only for display purposes: the phases were calculated without any alignment or rescaling. The right axis is for the phase residuals in cycles. The dashed line passes through the average of the phase residuals for the different bursts in equispaced bins and the colour contours indicate the distribution of the data as given by the kernel density estimation in Equation~\ref{equ:totprob} using the kernels from Equation~\ref{equ:kernel}. The best S/N bursts displayed on the left panels are 3, 4, 11, 12, 21, 27, 28, while the sufficient S/N bursts on the right are all except those with low counts (the low count ones are  1, 2, 6, 7, 8, 9, 14 and 19). In the right plot the trend is slightly washed out by the poorer signal to noise in some of the bursts and by the fact that the bins of the 20 second analysis do not encompass the rise and peak in the exact same way. A further effect is due to the rise and decay timescales which are slightly different among the bursts, since the drift and pullback follow these timescales.}
\label{fig:overall}
\end{figure*}

It is important to stress that this pattern is seen basically in all bursts where there is a sufficient S/N of the BO phases, i.e., in 20 out of 28 bursts (we also consider burst 28 even if it shows signs of photospheric radius expansion; see~\citealt{gal08b}). In Figure~\ref{fig:overall} we show this by overplotting two groups of bursts: the left panel shows the ones with the best signal to noise, while the right panel shows all the 20 bursts with S/N above the threshold (see caption for the IDs of the bursts). The bursts are aligned by their peaks in the 1 second lightcurves. We use two ways to show the trend of the phase residuals of the BOs with respect to the reference phase for each burst. First, a black dashed line passes through the weighted averages of these residuals in equispaced bins. Second, the colour contours in the background display the distribution of the detections obtained by applying a kernel density estimation on the data. This non-parametric way of displaying the data eliminates the bias one may introduce by fitting any trend or using any binning of the data. It is also mostly insensitive to the kernel used, which in our case is
\begin{equation}
P_i(t, \phi; t_i, \phi_i) = \frac{\textrm{box}(t; t_i, 20s)}{20} \frac{e^{-(\phi - \phi_i)^2 / 2 \sigma_i^2}}{\sigma_i \sqrt{2\pi}}
\label{equ:kernel}
\end{equation}
where $\textrm{box}(t; t_i, 20s)$ is the step function centered on $t_i$ with total width 20 seconds, the size of the bins, and $\sigma_i$ is the error on the phase residual $\phi_i$. The total probability is then given by
\begin{equation}
P(t, \phi) = \frac{\sum_i P_i(t, \phi; t_i, \phi_i)}{N}
\label{equ:totprob}
\end{equation}
where $N$ is the number of data points used. The trend is clearly visible in both panels. When we plot the 20 bursts together, the effect is slightly less visible because the phase residuals also include lower S/N burst oscillations. Furthermore, we have aligned the bursts by their peaks, but we use 20 second bins for the analysis, therefore the real (unbinned) peak location does not always correspond to the center of the bin so that its weight on the phase determination is different for each burst. Nonetheless, the fact that the trend is still visible further confirms our conclusions.

There is also another puzzling circumstance to notice: the overall movement of the BO phases is such that, when one considers the average burst oscillation phase over the entire burst length (from rise to tail), it is coincident with the APP pre/post burst phase within a few degrees. This phase coincidence is even more remarkable when considering the average BO phase (average across all bursts), that matches within $3^{\circ}$ the average APP phase, despite, per burst, the observed BO phase moves by about 0.05 cycles around the mean point. This behaviour is consistently observed throughout the outburst.

At longer timescales (${\sim}$ few hours to days), the APPs show the well known timing noise fluctuations tightly anti-correlated with the X-ray flux of the outburst~\citep{pat09f,has11}. Timescales shorter than a few hundred seconds are unexplored for APPs, mostly because the S/N is too small. Even if we have demonstrated above that the contamination of the APP on the BOs is negligible, it is legitimate to ask whether the drifts seen in the BOs, at timescales of few tens of seconds, are simply a consequence of an underlying fluctuation of the APPs at the same timescales, since APPs and BOs are strongly phase locked. However, we argue that this cannot be the case and the fluctuations of the BOs are truly different from those of the APPs. Indeed, since the start time of a burst is unrelated to the APP phases, we would expect a random initial position for the BOs as well. This is, however, not observed. The BOs are seen with a definite offset of ${\approx}0.05$ cycles, always in coincidence with the burst rise. Therefore this is a systematic effect unrelated to any residual undetected fluctuation of the APPs. 

It is worth noticing that the BOs have a known phase lag with respect to the APPs due to the different energy spectra of the thermonuclear bursts and accretion powered emission~\citep{wat06, wat08}. The magnitude of the effect reported here is of the order of 0.1 cycle, which corresponds to a time lag of about $300$ $\mu$s. The phase lag observed by \citep{wat06} is instead approximately an order of magnitude smaller than the effect we observe. Therefore what we are reporting here is not given by a change in the energy spectrum of the burst relative to the pre-burst spectrum. As a sanity check, we also tried several different energy selections to avoid that systematic effects due to the energy dependence might alter the results, but we see the same behaviour in all energy bands considered (with an overall decrease of the sensitivity due to the decreased counting statistics). 

When looking at the fractional amplitude of the BOs, the fundamental remains relatively constant (with amplitude of around 10-20\%), but sometimes there are brief changes in fractional amplitude, up to a factor of about 2. The second harmonic is detected only in a few pulse profiles, but when it is, it shows a systematic increase over time by a factor of about 2. Burst 28 is different, as noted for example in \citet{str03,wat05}, because the fractional amplitude of the BOs is $\sim 5$\%.

Overall the kinematics of the burst oscillation phases can be described as follows: 

\begin{itemize}

\item The BO phases start around the APPs. The initial leading observed in the early phases of some bursts is only tentatively detected and could be just the effect of APP wandering due to timing noise.

\item The BO phases suddenly drift ending up lagging the APPs  by about 0.05 - 0.1 cycles. 

\item The BO phases move back to the approximate location of the APPs in the next 100--200 seconds (pullback). 

\item Despite the drift and pullback, the burst oscillations phases somehow end up coincident with the APP phases within $3^{\circ}$, when averaging over all BO and APP phases (for a 10 km radius neutron star, this would correspond to a distance of less than 500 m).

\item Whatever mechanism produces the BOs, it must be such that it can shift around the APP emission location without being completely locked to it ($\lesssim\,0.15$ cycles) on short timescales (i.e., less than the burst duration), but it must be completely locked to it ($\lesssim 3^{\circ}$) on average, when considering the entire burst sequence.

\item The drift and pullback last for about 100-300 seconds. This is seen in basically all bursts, with little variation. 

\item The initial BO drift corresponds to a difference in frequency of the order of $10^{-3}$ Hz with respect to the accretion powered pulse frequency. This is the reason why this effect has not been consistently seen before, i.e. when using power spectra the sensitivity is below the threshold required to detect this phenomenon. 

\item The drift is seen at the very moment the burst begins, so the BOs must be produced (or have a centroid) close to the accretion hot spot/column, but not exactly matching its location.

\end{itemize}

\section{Discussion}
\label{sec:discussion}

First we describe what the observations mean in terms of the position of the centroid of the brighter region responsible for the BOs and then we discuss which mechanism could be at work. If the leading phase of the BOs observed at the beginning of some bursts is real, it would imply a centroid that is \emph{ahead} of the APP centroid, otherwise, the initial BO centroid is roughly coincident with the APP one. The sudden drift and lagging phase implies that the centroid of the BOs goes very quickly \emph{behind} the centroid of the APPs. The pullback phase implies that the BO source changes position moving back to a location slightly ahead of the APPs. The plateau phase implies that, in the last stages, the BO centroid adjusts to the original position of the APPs. A striking feature is that this evolution somehow \emph{knows} about the burst stage at which it takes place, because each step in the BO evolution happens consistently at the same burst stage for each burst.

We start our discussion assuming that the origin of the APPs can be identified with the accretion hot-spot, or more correctly with a shock in the accretion column close to the surface, but above the burning layer (e.g., \citealt{pou03, gie05, sul18}). This is the reference position for the phase. Also, note that 1) the burst flame has to cover the whole star, since the rms of the BOs is only $\sim 15$\% (and would otherwise be close to $100$\%); 2) we cannot explain the BOs by being produced by a moving accretion shock since that would make the BO rms drop by a factor of $\sim 10$ (i.e., the ratio between the flux at the top of the bursts and the persistent flux), bringing it down to 1--2\%, which is not observed.

Given the phase coincidence of the average BO phase with the APPs, the location where the former are generated is probably associated with the magnetic field \citep{Lovelace07,wat08,cav11} and it is probably safe to assume that, at least for this AMXP, ignition starts near the foot of the accretion column (we have no proof for this, but some tentative indications that this might be the case are presented by \citealt{good21}). If the accretion column is bent backwards with respect to the rotation of the star (in the sense that the foot of the accretion comes into view before the accretion shock), the ignition point occurring every time at the same location can explain the first feature: the fact that the BOs always start near to (or slightly leading\footnote{We stress again that this early leading is only tentatively detected in some bursts and should be verified with better data.}) the APPs. The sudden drift, with the BOs lagging the APPs, and the pullback, which brings the BOs back to the same phase where they started from, is harder to explain.

One of the best candidate model for the BOs is ocean oscillatory modes that lead to hotter and colder patches on the surface. Most of the modes discussed in the literature \citep{Heyl04,Piro05a,Cumming05,Berk08} neglect the magnetic field and drift \emph{against} the rotation of the star. The drifting velocity slows down as the star cools. This behaviour would fit with the lag of the phase and the fact that the lag decreases with time, but it fails to explain the fact that the BOs end up leading again the APPs at the end of the pullback. Also, while the BO frequency is slightly smaller than the spin frequency during the initial drift, it becomes slightly larger during the pullback.
However, it is difficult to find modes with a frequency of the drift of just $10^{-3}$ Hz (see Section \ref{sec:results} and the discussion in \citealt{cav11}). \citet{Heng09} discussed ocean modes including magnetic fields and even found modes that drift in the same direction of the rotation, which in principle could explain the leading of the BO phases. However, these modes cannot reproduce the lag. Furthermore, the drift velocity of these modes increases with cooling, which would lead to larger $\Delta\varphi_m$ expected near the end of the bursts, in contrast with what is observed in the plateau phase. Given that no mode mechanism is able to explain all the features we observe we discard this mechanism as an explanation for the BOs in XTE J1814-338.

Another possibility for the origin of the emission pattern is the creation of cyclonic and anticyclonic whirls in the burning ocean on top of a global circulation pattern \citep{spi02,cav19}. These structures do not seem to provide a viable explanation either, for reasons similar to the ones outlined above for the modes. Note also that none of these works includes the magnetic field and this already taints their applicability to our source. Furthermore, the structures that \citeauthor{cav19} observed in their simulations drift along the rotation of the star during the rise, in contradiction with the lag we observe after the sudden drift, and the ones expected in the decay phase of the bursts should always drift against the rotation of the star, in contrast with the BOs leading the APPs at the end of the pullback.

We speculate on a different explanation, which seems more compatible with the observations: the main features of the BOs could be explained if the burning fluid expands preferentially into a backward tilted accretion column. This picture is similar to what \citeauthor{Lovelace07} proposed in 2007 for the BOs in SAX J1808.4--3658. They suggested that, during the initial phases of a thermonuclear burst, an expanding fuel would be trapped by the magnetic field and initially drift backwards due to conservation of angular momentum.
This would be compatible with the BOs lagging the APPs during the burst rise.
Damped oscillations of the trapped cooling fluid due to magnetic torques would make the phase of the BOs oscillate around the one of the APPs. Combined with the ignition at the foot of the accretion column, this mechanism matches qualitatively well with the behaviour we see, but requires a degree of fine tuning. In this scenario the damping timescale of the oscillations would have to match exactly the cooling timescale of the bursts, which requires an improbable fine tuning, especially taking into account the difference in time scales between the 28 bursts, so that the fine-tuning needs to repeat for each of the 28 bursts.

On the other hand, the expansion of the burning fluid into the accretion column does not require fine tuning.  As we said at the beginning, ignition at the foot of a back tilted accretion column would explain the phase coincidence (or slight lead) at the very onset of the bursts. If during the rise of the bursts the flame heats up the surface, but \emph{at the same time} the fluid expands preferentially into the accretion column occupying a long enough region and lifting the accretion shock, thus moving it backwards thanks also to the effect of the burst flux on the column structure, the average projected position of the bright patch generating the BOs would quickly move back enough with respect to the original APP/shock position to explain the sudden drift\footnote{Why the expansion should take place preferably along the accretion column could be explained by the fact that at the magnetic pole the vertical  motion would be along the field lines, and thus more favourable, while at the magnetic equator the filed lines would restrain the same vertical lift. A caveat for this mechanism is the need to make the hot material in the column more visible.}. High speed Alfven waves could provide quick coupling and keep the light fluid in rigid rotation with the star \citep[see][for a discussion of various coupling mechanisms]{Cumming00}. The expansion expected during normal bursts is of order a few tens of meters \citep{Cumming00}, so depending on the angle of the magnetic column this could be a small change in projected position, but if the fluid is squeezed towards the accretion column the vertical expansion could be higher. At any rate, \citealt{lam09} showed that even a moderate change in position can lead to several tens of percent of cycle change in the phase, depending on the latitude of the centroid, as we observe. Changes of relativistic effects due to the height variation should not be relevant for the phase shift, because tens of meter are negligible compared to the star radius, while changes due to a variation of the centroid latitude could contribute to the shift \citep{lam09}.

The subsequent cooling of the hot fluid and decreasing of the burst flux would move the centroid of the BOs back to the foot of the accretion column. This naturally explains three key features we observe: the pullback phase, the BOs leading again the APPs at the end of it and the fact that this process is always tied to the burst profile. The plateau phase could be explained  by the accretion shock adjusting around its original position with any residual burst emission dwindling away. 
If one takes this description to face value, then the phases should follow the flux one to one: two moments, in the rise and in the tail, with the same flux, should have the same phase. That this is not the case is apparent from Figs. \ref{fig:bos27} and \ref{fig:bos11}. However, it is reasonable to expect that the flame would also perturb the accretion column, so that a completely symmetric situation during the rise and the tail is very unlikely \citep[see for example][and references therein, who discuss the possible effects of bursts on accretion, although not in magnetically dominated cases]{worpel2013,bhatta2018}.

\begin{figure}
\centering
\includegraphics[width=0.475\textwidth]{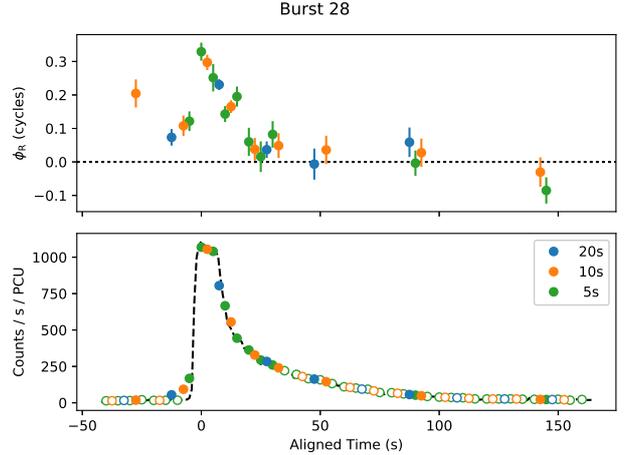}
\caption{Burst nr. 28. Similar to Figure~\ref{fig:bos27}, but here we just plot the lightcurve in the bottom panel and the phase residuals in the upper panel. Since the flux is higher, we can plot the results for bins of 5 seconds, 10 seconds and 20 seconds (along with the 1 second profile for the lightcurve). Again, empty circles in the lightcurves indicate that the BOs where not detected significantly in that bin. The most notable feature is that the trend we describe is present, but the amplitude of the drift is higher. This supports the idea presented in the Discussion about the burst flux affecting the column.} 
\label{fig:PRE}
\end{figure}

Accurate calculations beyond the scope of this paper are necessary to test whether this picture is correct or not. However, we could mention a further proof. We also analysed burst 28, the brightest one which may be a photospheric radius expansion burst: the results are plotted in Figure~\ref{fig:PRE}. First of all this burst displays the same trend as the others, but the most notable difference is that the \emph{amplitude} of the drift is higher, correlating with its higher intensity. This behaviour fits very nicely in our interpretation, since the fact that the burst shows brighter emission (by a factor of $\sim 2$) can easily lead to a stronger effect on the accretion column during the rise. The same behaviour was observed by \citet{str03} and \citet{wat05} for bursts 12 and 28 (note that \citet{str03} does not count the first burst so their counting is lower by one), where they noted a slower BO frequency. In our picture, the frequency of the BOs can be thought of as changing since the effective centroid of the BOs moves backwards leading to the impression of a slower rotation, while during the pullback the BO frequency would look as accelerating. Since the effect on the column is higher for the brighter bursts, it was apparent already in previous analysis, while to see it in weaker bursts the higher sensitivity of our 20 second timing analysis was needed. The fact that this evolution is barely detectable in some of the lowest count bursts also agrees with the effect of the burst flux on the column.

Finally, the fact that this discussion has several traits in common with that in~\citet{cav11}, which deals with the source IGR J17480-2446 -- a mildly recycled accreting pulsar, is suggestive that, with better data, a similar behaviour may be observed in all sources where the magnetic field is strong enough to show phase locking and what may have looked as small frequency fluctuations tentatively observed in the bursts of other AMXPs such as IGR J17511--3057, although deemed as statistically not significant \citep{alt10}, may turn out to be a similar combination of drift, pullback and plateau stages.

\section*{Data availability statement}

All the data used in this paper are available from the archives of \textit{RXTE}/NASA.

\section*{Acknowledgements}
AP acknowledges support from a Ramon y Cajal fellowship (RYC-2017-21810).

\bibliographystyle{mnras}
\bibliography{ms}

\label{lastpage}

\end{document}